%------------------------------------
%                     MACROS
%------------------------------------
%this is pmt20.sty, version August 1998.
%by F. J. Yndur\'ain
%Universidad Aut\'onoma de Madrid,
%Canto Blanco, 28049-Madrid.
%e-mail: fjy@delta.ft.uam.es
%comments welcome.

%---------------------------------------
 %text italic

%--------------------------------------------------------------------
%Fonts for authors, in references

 %caps and small caps
%---------------------------------------------------------------------

%----------------------------------------------------------
%Springer \tens, \bvec, \petit fonts.

      \font \ninebf                 = cmbx9
      \font \ninei                  = cmmi9
      \font \nineit                 = cmti9
      \font \ninerm                 = cmr9
      \font \ninesans               = cmss10 at 9pt
      \font \ninesl                 = cmsl9
      \font \ninesy                 = cmsy9
      \font \ninett                 = cmtt9
      \font \fivesans               = cmss10 at 5pt
						\font \sevensans              = cmss10 at 7pt
      \font \sixbf                  = cmbx6
      \font \sixi                   = cmmi6
      \font \sixrm                  = cmr6
						\font \sixsans                = cmss10 at 6pt
      \font \sixsy                  = cmsy6
      \font \tams                   = cmmib10
      \font \tamss                  = cmmib10 scaled 700
						\font \tensans                = cmss10
    
%---------------------------------------------------------------
  % petit-fonts
      \skewchar\ninei='177 \skewchar\sixi='177
      \skewchar\ninesy='60 \skewchar\sixsy='60
      \hyphenchar\ninett=-1
      \def\newline{\hfil\break}%
 %-----------------------------------------------------------
      \catcode`@=11
      \def\folio{\ifnum\pageno<\z@
      \uppercase\expandafter{\romannumeral-\pageno}%
      \else\number\pageno \fi}
      \catcode`@=12 % at signs are no longer letters

 %--------------------------------------------------------
      \newfam\sansfam
      \textfont\sansfam=\tensans\scriptfont\sansfam=\sevensans
      \scriptscriptfont\sansfam=\fivesans
      \def\sans{\fam\sansfam\tensans}
 %-----------------------------------------------------------

%-----------------------------------------------------------

      \def\petit{\def\rm{\fam0\ninerm}%
      \textfont0=\ninerm \scriptfont0=
\sixrm \scriptscriptfont0=\fiverm
       \textfont1=\ninei \scriptfont1=
\sixi \scriptscriptfont1=\fivei
       \textfont2=\ninesy \scriptfont2=
\sixsy \scriptscriptfont2=\fivesy
       \def\it{\fam\itfam\nineit}%
       \textfont\itfam=\nineit
       \def\sl{\fam\slfam\ninesl}%
       \textfont\slfam=\ninesl
       \def\bf{\fam\bffam\ninebf}%
       \textfont\bffam=\ninebf \scriptfont\bffam=\sixbf
       \scriptscriptfont\bffam=\fivebf
       \def\sans{\fam\sansfam\ninesans}%
       \textfont\sansfam=\ninesans \scriptfont\sansfam=\sixsans
       \scriptscriptfont\sansfam=\fivesans
       \def\tt{\fam\ttfam\ninett}%
       \textfont\ttfam=\ninett
       \normalbaselineskip=11pt
       \setbox\strutbox=\hbox{\vrule height7pt depth2pt width0pt}%
       \normalbaselines\rm

%------------------------------------------------------

      \def\bvec##1{{\textfont1=\tbms\scriptfont1=\tbmss
      \textfont0=\ninebf\scriptfont0=\sixbf
      \mathchoice{\hbox{$\displaystyle##1$}}{\hbox{$\textstyle##1$}}
      {\hbox{$\scriptstyle##1$}}{\hbox{$\scriptscriptstyle##1$}}}}}

%-----------------------------------------------------------

%-----------------------------------
.

					\mathchardef\gammav="0100
     \mathchardef\deltav="0101
     \mathchardef\thetav="0102
     \mathchardef\lambdav="0103
     \mathchardef\xiv="0104
     \mathchardef\piv="0105
     \mathchardef\sigmav="0106
     \mathchardef\upsilonv="0107
     \mathchardef\phiv="0108
     \mathchardef\psiv="0109
     \mathchardef\omegav="010A

%"versal", or slanted upper case greek characters

%--------------------------------
%The same characters are obtained with the following  definitions:
					\mathchardef\Gammav="0100
     \mathchardef\Deltav="0101
     \mathchardef\Thetav="0102
     \mathchardef\Lambdav="0103
     \mathchardef\Xiv="0104
     \mathchardef\Piv="0105
     \mathchardef\Sigmav="0106
     \mathchardef\Upsilonv="0107
     \mathchardef\Phiv="0108
     \mathchardef\Psiv="0109
     \mathchardef\Omegav="010A

%"versal", or slanted upper case greek characters

%---------------------------------------

\font\grbfivefm=cmbx5
\font\grbsevenfm=cmbx7
\font\grbtenfm=cmbx10 %for the greek bf family
\newfam\grbfam
\textfont\grbfam=\grbtenfm
\scriptfont\grbfam=\grbsevenfm
\scriptscriptfont\grbfam=\grbfivefm

\font\calbfivefm=cmbsy10 at 5pt
\font\calbsevenfm=cmbsy10 at 7pt
\font\calbtenfm=cmbsy10 %for the cal bf family
\newfam\calbfam
\textfont\calbfam=\calbtenfm
\scriptfont\calbfam=\calbsevenfm
\scriptscriptfont\calbfam=\calbfivefm

%boldface for upper case upright greek characters (\grbf)
% and upper case \cal characters (\calbf)

%-----------------------------------

      \def\bvec#1{{\textfont1=\tams\scriptfont1=\tamss
      \textfont0=\tenbf\scriptfont0=\sevenbf
      \mathchoice{\hbox{$\displaystyle#1$}}{\hbox{$\textstyle#1$}}
      {\hbox{$\scriptstyle#1$}}{\hbox{$\scriptscriptstyle#1$}}}}

%boldface for slanted latin and slanted greek characters. The 
%notation \rvec, reserved for arrow over character; see below
%------------------------------------

%-----------------------------------------------------------

\def\pmbf#1{\leavevmode\setbox0=\hbox{#1}%
\kern-.02em\copy0\kern-\wd0
\kern.04em\copy0\kern-\wd0
\kern-.02em\copy0\kern-\wd0
\kern-.03em\copy0\kern-\wd0
\kern.06em\box0 }

%"poor man" boldfaces, with automatic math scaling.

%--------------------------------------------------
%TEXT MACROS
%---------------------------------------------------------
						%time display with command \timestamp

						\def\monthname{%
   			\ifcase\month
      \or Jan\or Feb\or Mar\or Apr\or May\or Jun%
      \or Jul\or Aug\or Sep\or Oct\or Nov\or Dec%
   			\fi
							}%
					\def\timestring{\begingroup
   		\count0 = \time
   		\divide\count0 by 60
   		\count2 = \count0   % The hour, from zero to 23.
   		\count4 = \time
   		\multiply\count0 by 60
   		\advance\count4 by -\count0   % The minute, from zero to 59.
   		\ifnum\count4<10
     \toks1 = {0}%
   		\else
     \toks1 = {}%
   		\fi
   		\ifnum\count2<12
      \toks0 = {a.m.}%
   		\else
      \toks0 = {p.m.}%
      \advance\count2 by -12
   		\fi
   		\ifnum\count2=0
      \count2 = 12
   		\fi
   		\number\count2:\the\toks1 \number\count4 \thinspace \the\toks0
					\endgroup}%

				\newskip\abovelistskip      \abovelistskip = .5\baselineskip 
				\newskip\interitemskip      \interitemskip = 0pt
				\newskip\belowlistskip      \belowlistskip = .5\baselineskip
				\newdimen\listleftindent    \listleftindent = 0pt
				\newdimen\listrightindent   \listrightindent = 0pt

				%

%-------------------------------------------------------------------------

\def\petit{\def\rm{\fam0\ninerm}%
\textfont0=\ninerm \scriptfont0=\sixrm \scriptscriptfont0=\fiverm
\textfont1=\ninei \scriptfont1=\sixi \scriptscriptfont1=\fivei
\textfont2=\ninesy \scriptfont2=\sixsy \scriptscriptfont2=\fivesy
       \def\it{\fam\itfam\nineit}%
       \textfont\itfam=\nineit
       \def\sl{\fam\slfam\ninesl}%
       \textfont\slfam=\ninesl
       \def\bf{\fam\bffam\ninebf}%
       \textfont\bffam=\ninebf \scriptfont\bffam=\sixbf
       \scriptscriptfont\bffam=\fivebf
       \def\sans{\fam\sansfam\ninesans}%
       \textfont\sansfam=\ninesans \scriptfont\sansfam=\sixsans
       \scriptscriptfont\sansfam=\fivesans
       \def\tt{\fam\ttfam\ninett}%
       \textfont\ttfam=\ninett
       \normalbaselineskip=11pt
       \setbox\strutbox=\hbox{\vrule height7pt depth2pt width0pt}%
       \normalbaselines\rm
      \def\vec##1{{\textfont1=\tbms\scriptfont1=\tbmss
      \textfont0=\ninebf\scriptfont0=\sixbf
      \mathchoice{\hbox{$\displaystyle##1$}}{\hbox{$\textstyle##1$}}
      {\hbox{$\scriptstyle##1$}}{\hbox{$\scriptscriptstyle##1$}}}}}
%--------------------------------------------------------------------
% footnotes macros:					

      \def\footnoterule{\kern-3pt\hrule width 2cm\kern2.6pt}
      \newdimen\oldparindent\oldparindent=1.5em
      \parindent=1.5em
 
\newcount\footcount \footcount=0
\def\advftncnt{\advance\footcount by1\global\footcount=\footcount}
      % automatically numbered footnotes, in petit
      \def\fnote#1{\advftncnt$^{\the\footcount}$\begingroup\petit
      \parfillskip=0pt plus 1fil
      \def\textindent##1{\hangindent0.5\oldparindent\noindent\hbox
      to0.5\oldparindent{##1\hss}\ignorespaces}%
 \vfootnote{$^{\the\footcount}$}
{#1\nullbox{0mm}{2mm}{0mm}\vskip-9.69pt}\endgroup}
 %-------------------------------------------------------------------

 %------------------------------------------------------------------- 				

      \def\item#1{\par\noindent
      \hangindent6.5 mm\hangafter=0
      \llap{#1\enspace}\ignorespaces}
%-------------------------------------------------------------------
      
%note the difference with the TeXtbook \item, \itemitem, p. 355
%--------------------------------------------------------------------
      \def\leaderfill{\kern0.5em\leaders
\hbox to 0.5em{\hss.\hss}\hfill\kern
      0.5em}
%-----------------------------------------------------------------------
						\def\hb{\hfill\break}
%-----------------------------------------------------------------------

%dots in text
    \def\centerrule#1{\centerline{\kern#1\hrulefill\kern#1}}
%a rule centered, with margins equal to #1

%--------------------------------------------------------------------
%boxing it:

      \def\boxit#1{\vbox{\hrule\hbox{\vrule\kern3pt
						\vbox{\kern3pt#1\kern3pt}\kern3pt\vrule}\hrule}}
      %puts a box around it

      \def\tightboxit#1{\vbox{\hrule\hbox{\vrule
						\vbox{#1}\vrule}\hrule}}
						%puts a tight box around it

      \def\looseboxit#1{\vbox{\hrule\hbox{\vrule\kern5pt
						\vbox{\kern5pt#1\kern5pt}\kern5pt\vrule}\hrule}}
      %puts a loose box around it

      \def\youboxit#1#2{\vbox{\hrule\hbox{\vrule\kern#2
						\vbox{\kern#2#1\kern#2}\kern#2\vrule}\hrule}}
      %puts a  box around #1 with margins specified by #2

%----------------------------------------------------------

%various boxes, and tiles:

			\def\whitetile#1#2#3{\setbox0=\null
			\ht0=#1 \dp0=#2\wd0=#3 \setbox1=
\hbox{\tightboxit{\box0}}\lower#2\box1}

%\nulltile is identical to \nullbox as described in 
% the \TeX book, p. 82 (cf. also p.312):
			\def\nullbox#1#2#3{\setbox0=\null
			\ht0=#1 \dp0=#2\wd0=#3\box0}

%---------------------------------------------------------------

%-------------------------------------------------------------------

%common abreviations

\def\fig{\leavevmode Fig.}

\def\equ{\leavevmode Eq.}

%numbered:
\def\equn#1{\ifmmode \eqno{\rm #1}\else \equ~#1\fi}

%-------------------------------------------------------------

%-------------------------------------------------------------------

\def\tev{\ifmmode \mathop{\rm TeV}\nolimits\else {\rm TeV}\fi}
\def\gev{\ifmmode \mathop{\rm GeV}\nolimits\else {\rm GeV}\fi}
\def\mev{\ifmmode \mathop{\rm MeV}\nolimits\else {\rm MeV}\fi}
\def\kev{\ifmmode \mathop{\rm keV}\nolimits\else {\rm keV}\fi}
\def\ev{\ifmmode \mathop{\rm eV}\nolimits\else {\rm eV}\fi}

\def\chidof{\ifmmode
\mathop\chi^2/{\rm d.o.f.}\else $\chi^2/{\rm d.o.f.}\null$\fi}

\def\msbar{\ifmmode
\mathop{\overline{\rm MS}}\else$\overline{\rm MS}$\null\fi}

%------------------------------------------------------------

\def\physmatex{P\kern-.14em\lower.5ex\hbox{\sevenrm H}ys
\kern -.35em \raise .6ex \hbox{{\sevenrm M}a}\kern -.15em
 T\kern-.1667em\lower.5ex\hbox{E}\kern-.125emX\null}%

%----------------------------------------------------------------
\def\ref#1{$^{[#1]}$\relax}
%references, in square brackets
%------------------------------------------------------------------
%Journals

%-------------------------------------------------------------
% MATHEMATICAL MACROS

%(vector) arrow over character.

%allows stacking of 2 over 1.

%this allows to write under or above, the equality sign,
% with an underscore (_) order.

%the same with the "simeq" sign.

%the same with the "congruent" sign.

%the same under or above an arrow.

%the same with an identity sign 

%the same under, or above a similarity sign.

%the same under, or above an approx sign.

%-----------------------------------------------------------

%---------------------------------------------

%-----------------------------------------------------------------
\def\ddal{\mathop{\vrule height 7pt depth0.2pt
\hbox{\vrule height 0.5pt depth0.2pt width 6.2pt}
\vrule height 7pt depth0.2pt width0.8pt
\kern-7.4pt\hbox{\vrule height 7pt depth-6.7pt width 7.pt}}}
\def\sdal{\mathop{\kern0.1pt\vrule height 4.9pt depth0.15pt
\hbox{\vrule height 0.3pt depth0.15pt width 4.6pt}
\vrule height 4.9pt depth0.15pt width0.7pt
\kern-5.7pt\hbox{\vrule height 4.9pt depth-4.7pt width 5.3pt}}}
\def\ssdal{\mathop{\kern0.1pt\vrule height 3.8pt depth0.1pt width0.2pt
\hbox{\vrule height 0.3pt depth0.1pt width 3.6pt}
\vrule height 3.8pt depth0.1pt width0.5pt
\kern-4.4pt\hbox{\vrule height 4pt depth-3.9pt width 4.2pt}}}

%this produces the d'Alembertian operator,
% with correct display, script and scriptscript sizes

%--------------------------------------

%------------------------------------

\mathchardef\lap='0001
%this produces the Laplacian (upright uppercase Delta)

%-------------------------------------------

\def\lsim{\mathop{\setbox0=\hbox{$\displaystyle 
\raise2.2pt\hbox{$\;<$}\kern-7.7pt\lower2.6pt\hbox{$\sim$}\;$}
\box0}}
\def\gsim{\mathop{\setbox0=\hbox{$\displaystyle 
\raise2.2pt\hbox{$\;>$}\kern-7.7pt\lower2.6pt\hbox{$\sim$}\;$}
\box0}}
%these two represent, respectively,
% "less than, or sim", and "biger than, or sim". 
%to write under these signs,
% use the following commands, with #1 whatever goes under.

\def\gsimsub#1{\mathord{\vtop to0pt{\ialign{##\crcr
$\hfil{{\mathop{\setbox0=\hbox{$\displaystyle 
\raise2.2pt\hbox{$\;>$}\kern-7.7pt\lower2.6pt\hbox{$\sim$}\;$}
\box0}}}\hfil$\crcr\noalign{\kern1.5pt\nointerlineskip}
$\hfil\scriptstyle{#1}{}\kern1.5pt\hfil$\crcr}\vss}}}

\def\lsimsub#1{\mathord{\vtop to0pt{\ialign{##\crcr
$\hfil\displaystyle{\mathop{\setbox0=\hbox{$\displaystyle 
\raise2.2pt\hbox{$\;<$}\kern-7.7pt\lower2.6pt\hbox{$\sim$}\;$}
\box0}}
\def\gsim{\mathop{\setbox0=\hbox{$\displaystyle 
\raise2.2pt\hbox{$\;>$}\kern-7.7pt\lower2.6pt\hbox{$\sim$}\;$}
\box0}}\hfil$\crcr\noalign{\kern1.5pt\nointerlineskip}
$\hfil\scriptstyle{#1}{}\kern1.5pt\hfil$\crcr}\vss}}}
%---------------------------------------------------------------

\def\dd{{\rm d}}
%i, d for sqrt(-1) and differential

%the number e, and Euler's gamma

%trace

%----------------------------------------------------------------

%square black box
%------------------------------------------------------------

\def\frac#1#2{{#1\over#2}}
\def\dfrac#1#2{{\displaystyle{#1\over#2}}}
\def\tfrac#1#2{{\textstyle{#1\over#2}}}
\def\ffrac#1#2{\leavevmode
   \kern.1em \raise .5ex \hbox{\the\scriptfont0 #1}%
   \kern-.1em $/$%
   \kern-.15em \lower .25ex \hbox{\the\scriptfont0 #2}%
}%
%various forms of fractions
%-------------------------------

%----------------------------------------------------
%SOME CONVENIENT, SIMPLE FORMATS FOR BOOKS, BROCHURES AND PREPRINTS
%----------------------------------------------------

% brochure/preprint with two-sided printing
%--------------
% numbering of pages at the bottom, centered:

\def\brochureb#1#2#3{\pageno#3
\headline={\ifodd\pageno{\rheadline}
\else\lheadline\fi}
\def\rheadline{\hfil -{#2}-\hfil}
\def\lheadline{\hfil-{#1}-\hfil}
\footline={\hss -- \number\pageno\ --\hss}
\voffset=2\baselineskip}

%------------------------------------
\def\nada{\phantom{M}\kern-1em}

%write this after the cover page of a brochure,
% or preprint, for TWO SIDED printing
%omit for ONE SIDED PRINTING

%to end a brochure/paper with an even-numbered page.
%if using \brochureendcover: 

%if not using \endbrochurecover: write \bookendchapter

%------------------------------------------
%-----------------------------------------
%For book chapters.
% numbering at bottom (leading page not numbered):

\def\chapterb#1#2#3{\pageno#3
\headline={\ifodd\pageno{\ifnum\pageno=#3\hfil\else\rheadline\fi}
\else\lheadline\fi}
\def\rheadline{\hfil -{#2}-\hfil}
\def\lheadline{\hfil-{#1}-\hfil}
\footline={\hss -- \number\pageno\ --\hss}
\voffset=2\baselineskip}

%numbering at top (leading page not numbered)

%-------------------------------------------------------
\def\bookendchapter{\ifodd\pageno
 \vfill\eject\footline={\hfill}\headline={\hfill}\null \vfill\eject
 \else\vfill\eject \fi}
%at the end of the chapter, to finish with an even-numbered page.

\def\obookendchapter{\ifodd\pageno\vfill\eject
 \else\vfill\eject\null \vfill\eject\fi}
%at the end of the chapter, to finish with an odd-numbered page.

%-------------------------------------
%---------------------------------------
% For books:

%------------------------------------------------------------

\def\booksubsection#1{
\setbox0=\vbox{\hsize=0.85\hsize\tolerance=400\raggedright\hfuzz=4mm
\noindent{\fib #1}\smallskip}\goodbreak\vskip0.45cm\box0
\nobreak
\noindent}
%--------------------------------------------------

%--------------------------------------------------
% For brochures/papers:

%--------------------------------------------

%---------------------------------------------------
% figure captions

\def\figurasc#1#2{\petit{\noindent\sc#1}\ #2}
%---------------------------------------------------

%captiontype
\def\captiontype{\tolerance=800\hfuzz=1mm\raggedright\noindent}

%-----------------------------------------------------

%abstracttype INCLUDES size of box (valid also for table of contents),  
%AND specification of \petit font.

\def\abstracttype#1{
\hsize0.7\hsize\tolerance=800\hfuzz=0.5mm \noindent{\fib #1}\par
\medskip\petit}

%--------------------------------------------------

%------------------------------------------------------
\def\hb{\hfill\break}

%FONTS

\font\twelverm=cmr12 %for titles
%for titles
%-------------------------------
 %for reference in quotations; for running head
 %for text in quotations
 %small caps
%----------------------------------------------------------------------
\font\fib=cmfib8

%for sections and chapter headings
%--------------------------------------------------------------------
%Fonts for authors, in references

\font\sc=cmcsc10 %caps and small caps
%--------------------------------------------------------

\font\addressfont=cmbxti10 at 9pt%for (professional) address

 %text italic

%----------------------------------------------------------------
\catcode`@=11 % borrow the private macros of PLAIN (with care)

\newdimen\pagewidth \newdimen\pageheight \newdimen\ruleht
 \maxdepth=2.2pt  \parindent=3pc
\pagewidth=\hsize \pageheight=\vsize \ruleht=.4pt
\abovedisplayskip=6pt plus 3pt minus 1pt
\belowdisplayskip=6pt plus 3pt minus 1pt
\abovedisplayshortskip=0pt plus 3pt
\belowdisplayshortskip=4pt plus 3pt

\newinsert\margin
\dimen\margin=\maxdimen
%\count\margin=0 \skip\margin=0pt % marginal inserts take up no space

%----------------------------------------------------------

%margins:\topmargin=  ,\bottommargin=  ,\leftmargin=  ,\rightmargin=
%\advancetopmargin=  ,\advancebottommargin=  ,etc

%%Care should be exercised in putting these instructions
% AFTER any \magnification!

\newdimen\paperheight \paperheight = \vsize
\def\topmargin{\afterassignment\@finishtopmargin \dimen0}%
\def\@finishtopmargin{%
  \dimen2 = \voffset		% Remember the old \voffset.
  \voffset = \dimen0 \advance\voffset by -1in
  \advance\dimen2 by -\voffset	% Compute the change in \voffset.
  \advance\vsize by \dimen2	% Change type area accordingly.
}%
\def\advancetopmargin{%
  \dimen0 = 0pt \afterassignment\@finishadvancetopmargin \advance\dimen0
}%
\def\@finishadvancetopmargin{%
  \advance\voffset by \dimen0
  \advance\vsize by -\dimen0
}%
\def\bottommargin{\afterassignment\@finishbottommargin \dimen0}%
\def\@finishbottommargin{%
  \@computebottommargin		% Result in \dimen2.
  \advance\dimen2 by -\dimen0	% Compute the change in the bottom margin.
  \advance\vsize by \dimen2	% Change the type area.
}%
\def\advancebottommargin{%
  \dimen0 = 0pt\afterassignment\@finishadvancebottommargin \advance\dimen0
}%
\def\@finishadvancebottommargin{%
  \advance\vsize by -\dimen0
}%
\def\@computebottommargin{%
  \dimen2 = \paperheight	% The total paper size.
  \advance\dimen2 by -\vsize	% Less the text size.
  \advance\dimen2 by -\voffset	% Less the offset at the top.
  \advance\dimen2 by -1in	% Less the default offset.
}%
\newdimen\paperwidth \paperwidth = \hsize
\def\leftmargin{\afterassignment\@finishleftmargin \dimen0}%
\def\@finishleftmargin{%
  \dimen2 = \hoffset		% Remember the old \hoffset.
  \hoffset = \dimen0 \advance\hoffset by -1in
  \advance\dimen2 by -\hoffset	% Compute the change in \hoffset.
  \advance\hsize by \dimen2	% Change type area accordingly.
}%
\def\advanceleftmargin{%
  \dimen0 = 0pt \afterassignment\@finishadvanceleftmargin \advance\dimen0
}%
\def\@finishadvanceleftmargin{%
  \advance\hoffset by \dimen0
  \advance\hsize by -\dimen0
}%
\def\rightmargin{\afterassignment\@finishrightmargin \dimen0}%
\def\@finishrightmargin{%
  \@computerightmargin		% Result in \dimen2.
  \advance\dimen2 by -\dimen0	% Compute the change in the right margin.
  \advance\hsize by \dimen2	% Change the type area.
}%
\def\advancerightmargin{%
  \dimen0 = 0pt \afterassignment\@finishadvancerightmargin \advance\dimen0
}%
\def\@finishadvancerightmargin{%
  \advance\hsize by -\dimen0
}%
\def\@computerightmargin{%
  \dimen2 = \paperwidth		% The total paper size.
  \advance\dimen2 by -\hsize	% Less the text size.
  \advance\dimen2 by -\hoffset	% Less the offset at the left.
  \advance\dimen2 by -1in	% Less the default offset.
}%
%--------------------------------------------------------------------------

\def\onepageout#1{\shipout\vbox{ % here we define one page of output
    \offinterlineskip % butt the boxes together
    \vbox to 3pc{ % this part goes on top of the 44pc pages
      \iftitle % the next is used for title pages
        \global\titlefalse % reset the titlepage switch
        \setcornerrules % for camera alignment
      \else\ifodd\pageno \rightheadline\else\leftheadline\fi\fi
      \vfill} % this completes the \vbox to 3pc
    \vbox to \pageheight{
      \ifvoid\margin\else % marginal info is present
        \rlap{\kern31pc\vbox to\z@{\kern4pt\box\margin \vss}}\fi
      #1 % now insert the main information
      \ifvoid\footins\else % footnote info is present
        \vskip\skip\footins \kern-3pt
        \hrule height\ruleht width\pagewidth \kern-\ruleht \kern3pt
        \unvbox\footins\fi
      \boxmaxdepth=\maxdepth
      } % this completes the \vbox to \pageheight
    }
  \advancepageno}

\def\setcornerrules{\hbox to \pagewidth{\vrule width 1pc height\ruleht
    \hfil \vrule width 1pc}
  \hbox to \pagewidth{\llap{\sevenrm(page \folio)\kern1pc}%
    \vrule height1pc width\ruleht depth\z@
    \hfil \vrule width\ruleht depth\z@}}
\newbox\partialpage

%-------------------------------------
% Page layout

%%Care should be exercised in putting the instructions
% below AFTER any \magnification!
%-----------------------------------

%for arbitrary dimensions:

%----------------------------------

%------------------------------------
%                     TEXTO
%------------------------------------

\input epsf.sty
\raggedbottom
\footline={\hfill}
\rightline{FTUAM 01-06}
\rightline{hep-ph/}
\rightline{April, 2001}
\bigskip
\hrule height .3mm
\vskip.6cm
\centerline{{\twelverm Restriction on  SUSY Masses from  $Z\to \hbox{hadrons}$}}
\medskip
\centerrule{.7cm}
\vskip1cm

\setbox9=\vbox{\hsize65mm {\noindent\fib M. J. Herrero and F. J. 
Yndur\'ain} 
\vskip .1cm
\noindent{\addressfont Departamento de F\'{\i}sica Te\'orica, C-XI,\hb
 Universidad Aut\'onoma de Madrid,\hb
 Canto Blanco,\hb
E-28049, Madrid, Spain.}\hb}
\smallskip
\centerline{\box9}
\bigskip
\setbox0=\vbox{\abstracttype{Abstract}We remark that the precision of recent determinations of 
$\alpha_s(M^2_Z)$ is such that one can get bounds on supersymmetric partner
masses (squark and gluino) by requiring consistency of determinations of 
$\alpha_s$ at ``low" energies, where those particles 
do not contribute, and on the $Z$ peak. For approximately 
degenerate squarks and gluinos with mass $\widetilde{m}$, 
we find the bound $\widetilde{m}\geq 173$ GeV, at the $2\sigma$ level. 
At the  $2.5\sigma$ level, $\widetilde{m}\geq 121$ GeV.}
\centerline{\box0}
\vskip1truecm

\booksubsection{1 Introduction}
In the past years, determinations of the strong coupling 
at various energy scales have been performed 
with increasing degree of accuracy. 
In the table 1 we present a summary of those that 
have been pushed to the NNLO level, 
showing (as has become customary) the 
values of $\alpha_s$ extrapolated to the scale of the $Z$ mass.

\setbox0=\vbox{\petit
\medskip
\setbox1=\vbox{\offinterlineskip\hrule
\halign{
&\vrule#&\strut\hfil#\hfil&\quad\vrule\quad#&
\strut\quad#\quad&\quad\vrule#&\strut\quad#\cr
 height2mm&\omit&&\omit&&\omit&\cr 
& \kern.5em Process&&${\textstyle\hbox{Average}\;
 Q^2}\atop{\textstyle \hbox{or}\; Q^2\;\hbox{range}\;[\gev]^2}$&
& $\alpha_s(M_Z^2)$\kern.3em& \cr
 height1mm&\omit&&\omit&&\omit&\cr
\noalign{\hrule} 
height1mm&\omit&&\omit&&\omit&\cr
&\phantom{\Big|}  DIS $\nu$, Bj&&1.58&&$0.121^{+0.005}_{-0.009}$\phantom{l}& \cr
\noalign{\hrule}
&\phantom{\Big|}  DIS $\nu$, GLS&&3&&$0.112\pm0.010$\phantom{l}& \cr
\noalign{\hrule}
&\phantom{\Big|}  $\tau$ decays&&$(1.777)^2$&&$0.1181\pm0.0031$\phantom{l}&\cr
\noalign{\hrule}
&\phantom{\Big|}  $e^+e^-\to{\rm h}$&&$100 - 1600$&&$0.128\pm0.025$\phantom{l}&\cr
\noalign{\hrule}
&\phantom{\Big|} $e^+e^-\to$ h&&${2.5} - {230}$&&$0.123\pm0.007$\phantom{l}&\cr
\noalign{\hrule}
&\phantom{\Big|}  ($\nu\,N;\;xF_3$)&&$8 - 120$&&$0.1153\pm0.0041$\phantom{l}& \cr
\noalign{\hrule}
&\phantom{\Big|} $ep$&&${3.5} - {5000}$&&$0.1163\pm0.0014$\phantom{l}&\cr
\noalign{\hrule}
&\phantom{\Big|}  $Z\;{\rm width}$&&$(91.2)^2$&&$0.1230\pm0.0038$\phantom{l}&\cr
\noalign{\hrule}
&\phantom{\Big|}   GrandLEP, on $Z$&&$(91.2)^2$&&$0.1185\pm0.0030\phantom{l}$&\cr
 height1mm&\omit&&\omit&&\omit&\cr
\noalign{\hrule}}
\vskip.05cm}
\centerline{\box1}
\smallskip
\centerline{\petit Table 1}
\centerrule{6cm}
\medskip}
\box0
\noindent We explain the entries in this table. DIS $\nu$ means deep inelastic scattering by neutrinos, 
Bj stands for the Bjorken, and GLS for the Gross--Llewellyn Smith 
sum rules. The value coming from $\tau$ decays is also included.   The two entries $e^+e^-\to{\rm h}$ refer
to determinations  from  $e^+e^-$ annihilation to hadrons in PETRA and 
TRISTAN. All of these 
we have taken from the review of Bethke\ref{1}, where one 
can find references to original papers. The results labeled  $\nu\,N;\;xF_3$ 
and $ep$ are from the very recent analysis of ref.~2, which improve (and 
supersede) those quoted by Bethke.

Finally, the figures for the last two processes recorded in Table~1,
$$\eqalign{\alpha_s^{\petit (Z\;{\rm width})}(M_Z^2)=&0.1230\pm0.0038\cr
\alpha_s^{\petit (\hbox{GrandLEP, on $Z$})}(M_Z^2)=&0.1185\pm0.0030,\cr}\eqno{(1)} $$
we have taken the results of the LEP analyses, as updated in ref.~3, 
and corrected for the value $M_H=115\;\gev$ for the Higgs mass. 
The first value corresponds to the determination from 
the $Z$ hadronic width (to be more precise, from the ratio $R_l=\Gamma_{\rm had}/\Gamma_{\ell}$),
 and the
second  to that obtained with a fit to all LEP observables, on the $Z$ 
peak. We will however not use this last value here.
  
All determinations are compatible with 
one another, within errors. What is more, 
the precision of both the ``low energy" 
($Q^2\leq(70)^2\;\gev^2$) and the determination 
given by the decay $Z\to hadrons$ or  $e^+e^-\to$ hadrons on the $Z$  are, separately, such 
that we can use them to constrain effects of heavy 
particles (particles with mass larger than the $Z$ mass). 
In particular, here we will explore bounds on  squarks and gluino masses, that 
for simplicity we take of the same order of magnitude:
$m_{\tilde{q}}\sim m_{\tilde{g}}\sim \widetilde{m}$.

The strategy is as follows. 
Clearly, the values of the masses of SUSY particles has no influence on the determinations 
of $\alpha_s$ presented in the table above, {\sl except} 
on those based on $Z\to\hbox{hadrons}$, or  $e^+e^-\to$ hadrons on the $Z$,
  where the energy is much
higher than those of the other  processes. 
If we then evaluate the value of $\alpha_s$ obtained by averaging all 
determinations, {\sl excluding those obtained on the $Z$ peak}, we find 
$$\alpha_s^{(\hbox{\petit exc. $Z$ peak})}(M_Z^2)=0.116835\pm0.00120.\eqno{(2)}$$ 

The determination of $\alpha_s$ from $Z$ width showed in Table 1 
was made fixing the mass of the
Higgs particle  to 115 \gev, and assuming no SUSY partners ($\widetilde{m}=\infty$), 
and it is compatible with $\alpha_s^{(\hbox{\petit exc. $Z$ peak})}(M_Z^2)$, within errors. 
To be precise, we note that the value
 $\alpha_s^{(\hbox{\petit $Z$ width})}(M_Z^2)$ is slightly more
than  one sigma away from the average $\alpha_s^{(\hbox{\petit exc. $Z$ peak})}(M_Z^2)$; 
it is actually almost $1.5 \sigma$ off.  
(In calculating this, we add in quadrature the errors in both 
$\alpha_s^{(\hbox{\petit exc. $Z$ peak})}(M_Z^2)$ 
 and $\alpha_s^{(\hbox{\petit $Z$ width})}(M_Z^2)$
to get the effective error of 
$\pm0.0040$). Likewise, the determination from  $e^+e^-\to$ hadrons on the $Z$  
is compatible with lower energy determinations (half a sigma away only). 
Actually, both determinations on the $Z$ 
produce results slighly higher than those 
found at lower energies. Now, when one calculates the corrections due to a Higgs particle, 
with a mass $M_H> 115$ \gev, or  
SUSY contributions with particles of mass $\widetilde{m}$ (see below), they tend to 
increase the effective values of $\alpha_s$ on the $Z$ (for the Higgs, this occurs if 
increasing $M_H$). Therefore,     
 if we now vary $M_H$ and/or $\widetilde{m}$, the quantity 
$\alpha_s^{\petit (Z\;{\rm width})}(M_Z^2)$ will vary, separating even more from the 
low energy determination of $\alpha_s$: 
this will provide us with lower bounds on the SUSY masses.\fnote{Upper bounds on the Higgs mass 
would also follow with the same procedure, but the results do not improve the bounds 
 found with the standard method of fitting all observables; see e.g. ref.~3.}

\booksubsection{2 Bound on $\widetilde{m}$}
If squarks and gluinos existed with masses of the order of $\widetilde{m}$, 
then they would contribute to the 
decay $Z\to \hbox{hadrons}$ through 
diagrams like those in \fig~1. If we renormalize on the quark mass shell, then the 
contributions from the second diagram there 
cancel the divergence of the first.  
  If we include SUSY effects, 
we then get the decay rate, to first order in $\alpha_s$, and neglecting 
quark masses, 
$$\Gamma(Z\to\hbox{hadrons})=\Gamma^{(0)}\left\{1+
\left[1-K_{\rm SUSY}\right]\dfrac{\alpha_s}{\pi}\right\},\eqno{(3a)}$$
and we have
$$\eqalign{K_{\rm SUSY}=&\tfrac{1}{18}J(M_Z^2/\widetilde{m}^2)\dfrac{M_Z^2}{\widetilde{m}^2},\cr
J(\xi)=&\dfrac{12}{\xi}+\dfrac{24}{\xi^2}\int_0^1\dd x\;
\dfrac{[1-\xi x(1-x)]\log[1-\xi x(1-x)]}{1-x}\simeq 1+\tfrac{1}{15}\xi.\cr}.\eqno{(3b)}$$
$\Gamma^{(0)}$ is the partonic level decay width. 
We can interpret the value of $\alpha_s$ here as that obtained from 
lower energy determinations, that should not be 
contaminated by SUSY contributions; thus, we should take
$\alpha_s(M^2_Z)=\alpha_s^{(\hbox{\petit exc. $Z$ peak})}(M_Z^2)$, 
within allowed errors.

\topinsert{
\setbox0=\vbox{\epsfxsize 12.truecm\epsfbox{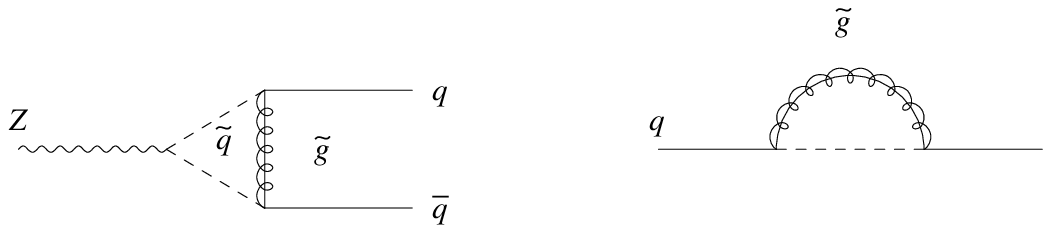}} 
\setbox6=\vbox{\hsize 10.truecm\captiontype\figurasc{Figure 1. }{Diagrams 
involving SUSY partners  
contributing to the decay  $Z\to \hbox{hadrons}$, and to renormalization 
of the wave function.}\hb
\vskip.1cm} 
\medskip
\centerline{\box0}
\centerline{\box 6}
\medskip
}\endinsert

On the other hand, the experimental figures given from $Z$ width  in  Table~1 are  
obtained neglecting possible SUSY contributions, so we  also have 
$$\Gamma(Z\to\hbox{hadrons})=\Gamma^{(0)}
\left\{1+\dfrac{\alpha_s^{(\hbox{\petit $Z$ width})}(M_Z^2)}{\pi}\right\}.\eqno{(4)}$$

Equating with the previous expression, we find
$$\alpha_s^{(\hbox{\petit $Z$ width})}(M_Z^2)=\alpha_s^{(\hbox{\petit exc. $Z$ peak})}(M_Z^2) 
\left\{1-\tfrac{1}{18}J(M_Z^2/\widetilde{m}^2)\dfrac{M_Z^2}{\widetilde{m}^2}\right\}.\eqno{(5)}$$
and a similar formula for $\alpha_s^{\petit (\hbox{GrandLEP, on $Z$})}(M_Z^2)$.

We may  fix the mass of the Higgs particle to its experimental 
value/lower bound, $M_H= 115\,\gev$ as increasing it would only improve the bounds. 
Taking thus the value of  
$\alpha_s^{(\hbox{\petit $Z$ width})}(M_Z^2)$
 as in Eq.~(1) above 
we find that, unless $\widetilde{m}$ is larger than 
a certain bound, we have, to satisfy Eq.~(5), to push
 $\alpha_s^{(\hbox{\petit exc. $Z$ peak})}(M_Z^2)$ in Eq.~(2)  beyond its 
allowed limits of variation. 
  
The bounds we obtain are then, 
$$\widetilde{m}\geq\cases{
173\;\gev,\quad 2\sigma\cr
121\;\gev,\quad 2.5\sigma.\cr}\eqno{(6)}$$

Note that, because the values of $\alpha_s^{(\hbox{\petit $Z$ width})}(M_Z^2)$ and 
$\alpha_s^{(\hbox{\petit exc. $Z$ peak})}(M_Z^2)$ differ by a little more than $1.5\sigma$, no bound can
be given at the $1.5\sigma$ level.

These bounds  are not very different from what one gets in studies 
with the Tevatron,\ref{4} but is perhaps more transparent and in
particular independent on assumptions about decay properties of SUSY 
partners, and about SUSY GUT relations on masses and/or couplings\ref{4}.

%--------------------------------
\vfill\eject
\booksubsection{Acknowledgments}

 The financial support of CICYT (Spain, proyecto \# FPA 2000-0980), is  gratefully acknowledged. 

\booksubsection{References}
\item{1.- }{S. Bethke, {\sl J. Phys.} {\bf G26} (2000) {R27}.}
\item{2.- }{J.~Santiago and F. J. Yndur\'ain,  FTUAM 01-01, 2001 (hep-ph/0102247).}
\item{3.- }{D. Strom, ``Electroweak measurements on the $Z$ resonance", 
Talk presented at the 5th Int. Symposium on Radiative Corrections, RadCor2000, 
Carmel, Ca., September 2000.  For discussion of the
Higgs, see also E. Tournefier, Talk at the Int. Seminar ``QUARKS '98",  Suzdal (hep-ex/9810042).}
\item{4.- }{Particle Data Group, Eur. Phys. Journal {\bf C15} (2000) 1.}
\item{}{} 

\bye